# Machine Learning Promoting Extreme Simplification of Spectroscopy Equipment

Jianchao Lee,[1]* Qiannan Duan,[1,2†] Sifan Bi,[1†] Ruen Luo,[1] Yachao Lian,[1] Hanqiang Liu,[3] Ruixing Tian,[1] Jiayuan Chen,[1] Guodong Ma,[1] Jinhong Gao[1], Zhaoyi Xu[2]

[1]Department of Environment Science, Shaanxi Normal University, Xi'an 710062, China
[2]School of the Environment, Nanjing University, Nanjing 210046, China
[3]Shool of Computer Science, Shaanxi Normal University, Xi'an 710062, China
*Corresponding author, e-mail: *jianchaolee@snnu.edu.cn*
†These authors contributed equally to this work

**Abstract:** The spectroscopy measurement is one of main pathways for exploring and understanding the nature. Today, it seems that racing artificial intelligence will remould its styles. The algorithms contained in huge neural networks are capable of substituting many of expensive and complex components of spectrum instruments. In this work, we presented a smart machine learning strategy on the measurement of absorbance curves, and also initially verified that an exceedingly-simplified equipment is sufficient to meet the needs for this strategy. Further, with its simplicity, the setup is expected to infiltrate into many scientific areas in versatile forms.

Absorption properties of light rays by matter are an important area of spectroscopy research [1]. At the same time, it is also an important means in many scientific fields. The spectral range mainly involves from X-rays of few nanometers to infrared light of thousands of micrometers [2]. Its ability to interpret and define the nature of matters has a profound effect on natural sciences and engineering technology. After nearly a hundred years of accumulation, researchers have established a huge knowledge system based on spectral data, and widely applied it in chemical [3, 4], biological [5, 6], medical [7] and other fields [8].

There are many techniques for obtaining various kinds of spectral data. The common methods start from splitting light beams to obtain spectral lines with an extremely-narrow wavelength distribution [2]. These spectral lines then work on the test samples, and the remaining energy of the rays is detected [8]. After dividing the interesting spectrum into hundreds, or even thousands of small segments, the split data is finally integrated to obtain a whole spectral curve [8, 9]. This route is adopted by many conventional spectroscopy instruments. The resolution of spectrum always is a focus of spectroscopic researches. The natural width of the spectrum sets a limit to the resolution of the spectroscopy. After the laser was applied in the spectroscopic study, it heavily stroked the technical bottleneck [8, 10]. We are conscious of that the complexity stems from the undeleted optical-splitting operations and destines the cost in the spectrum measurement [2].

The opposite direction to the optical splitting is a continuous spectrum band. For polychromatic light, especially for the light having complex wavelength spectrum, its intensity fluctuates distinctly along wavelength. When a matter absorbing a portion of the light, the residual spectrum is often considered to being still complex and difficult to be analyzed. In previous studies, these kinds of light radiations were not considered as proper information carriers for data acquisition [3]. But just in this state, a technology route without pursuing high resolution will fundamentally change the spectral instruments and dramatically reduce the cost of obtaining spectroscopy data. This route will drive the spectral measurement into some new forms, and penetrate into many details of natural science.

It should be pointed out that the complex spectrum band is also a carrier of matter properties. For example, a light beam containing complex spectral bands is maintained constant evenly, and then travels through a given solution. The residual spectrum of the beam is unique and constant even keeping complex. Following an appropriate pathway to understand the complex results, researchers can still know some spectral properties of the matter. Of course, the mining of information in such data also requires some ideas and methods for analyzing complex systems.

In recent years, machine learning (ML) has been developing rapidly [11]. In many areas including business and society management, the ML application has achieved a lot of meaningful innovation [11,12,13]. This is due mainly to some distinctive capabilities of ML, e.g., the outstanding performance on handling complex systems, which are hardly tamed on the base of existing knowledge [14,15]. The complex spectrum just is a typical one.

In the course of understanding complex systems, the algorithm of ML, especially deep learning (DL), mainly executes supervised training [16]. The basic strategy is to collect and obtain a big database that describes the research object, then to build a neural network consisting of numerous parameters (generally $>10^7$), later to raise the model with training on the obtained data [16,17]. In this process, the acquisition of a large amount of data is a key step. In today's torrent of artificial intelligence, there has been a good breakthrough in the algorithms and hardware for our research. However, most of spectrum technologies are costly and low-efficient in the acquisition of required big data. So we have to find a low-costs and simple carrier for spectrum measurement and open up an efficient access.

## Results
### Overall scheme
How to obtain speedily a large amount of spectral information is a challenging step in implementing ML. The complex ray without being split is a perfect carrier, which means we can design a spectral equipment without monochromators. Further, with being modulated under diverse filtering conditions, the original light will generate numerous light beams consisting of diverse spectra. Then these beams are combined into a two-dimensional form technically. For instance, with the aid of a 1000×1000 filter array (called as extremely-combined filter, ECF), the beam patterns will increase in the combinatorial number from 1 to $\sim 10^6$. The strategy is described vividly as, that numerous beams act on a test sample independently, and the residual of these beams are recorded in a combined result (Fig. 1a). It can be found that the best presentation of the result is an image form composed of colored pixels, in our lab called as combined-spectrum images (SCiM). The data contained in the image are normalized in two-dimensional matrices. After constructing a large dataset of these SCiMs, it is possible to train an equivalent DL model for understanding the research object. In addition, when the input is images, a kind of DL models, convolutional neural network (CNN) model is an exceedingly-powerful tool today.

In common spectral ranges, such as ultraviolet, visible, and infrared wavelengths, the SCiM is also qualified to take responsibility for data acquisition, after being improved in some operations. It can be expected that the obvious advantage is on simplifying the external form of the equipments.

For verifying and evaluating the proposed technical strategy, we focused on designing and improving the new equipment thoroughly. At last, we practiced its measurement ability in a small and complete hypothesis space that we can fulfill the job with a reasonable time cost. Following the hint of the ideas, we constructed a set of equipment (Fig. 1b). At the preliminary stage, we chose visible light as the physical example. More specifically, the experiments are carried out to evaluate the ML-predicting ability of the SCiM images to the absorbance curves of some samples. The sample group was limited to a small one so that we can collect enough data to cover an entire question space.

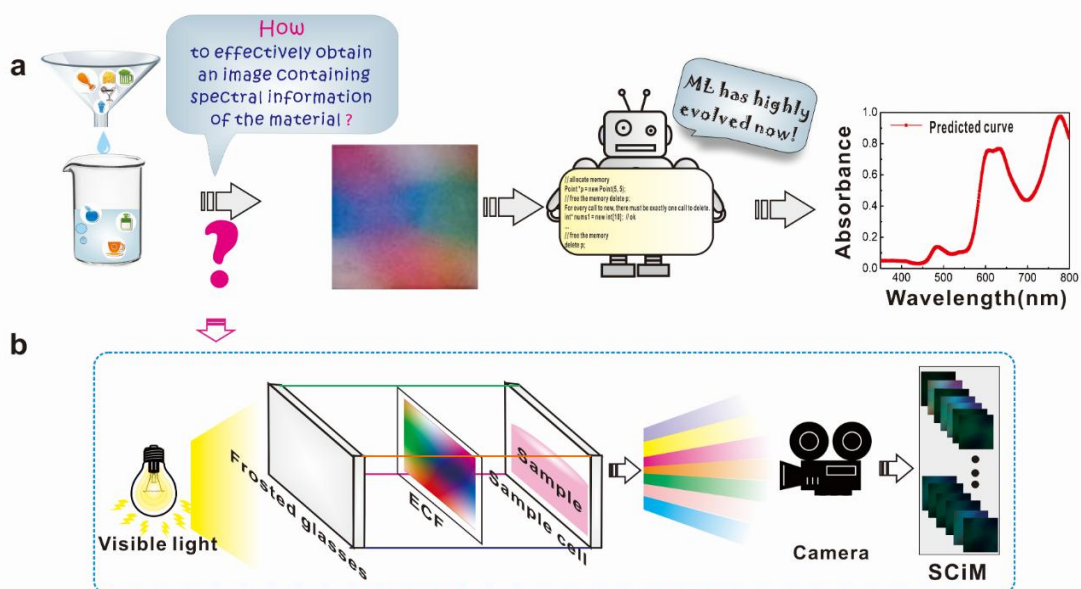

**Fig. 1. Schematic of the technical thought and its realization.** (**a**) The introduction of machine learning in the measurement of spectral properties of substances. In the whole technical route, there still are some barriers necessary to be broken. SCiM seems like a promising tool. (**b**) In the area of visible light, a complete setup for acquiring SCiMs. Mainly includes an original light source, a group of frosted-glass slides, an ECF, a sample cell and a camera device. Among them, the ECF is the key to producing a large amount of spectrum data. Order of magnitude in the ECF is about $10^5$.

**Producing SCiMs and absorbances**

A large number of SCiMs are the basis of building datasets for the next ML. The raw data of the SCiM image was obtained using the device constructed above (Fig. 1b). Here, we use a number of raw solutions (listed in Supplementary Table 1) to precisely prepare thousands of different combinations of sample solutions. This sampling technique will prompt the automatic data production in future. In the running of the device, an original visible light passed through an ECF to create strictly $\sim 10^5$ kinds of steadily-modulated light beams, and then these beams passed through a sample solution, and finally were captured as a raw image of SCiM by a camera (shooting parameters in Supplementary Table 2). Due to the difference of light-absorption properties, the captured SCiM for each sample will be specific. In this work, we acquired the SCiMs of thousands of solutions, and constructed SCiM sets (Fig. 2) for the later machine learning.

At a steady optical condition, the acquisition speed of SCiMs is determined by the shift time of the test samples. In current work, the time cost reached a level of ~20 seconds. With some aiding of proper automatic-sampling in future, the cost will be reduced to seconds, which means a production of $10^4$ SCiM images in a day. Considering the data amount ($10^5$) in a SCiM, the daily production of data is expected to reach a level of $10^9$.

The absorbance curve is the target of prediction and is also an indicator used to supervise the CNN learning. In conventional measurement techniques, the absorbance curve of a sample solution is obtained in a wavelength-scanning mode. The work of thousands of samples is time-consuming and difficult to maintain all samples in stability for long period. So we adopted combining measurement and weighted calculation to obtain enough data for absorbance curves. In the preliminary experiments, we measured the absorbance curves of several solutions and their mixtures by a UV-Vis spectrophotometer. The results show that the curves of the mixed solutions equal exactly to the weighted-sum curve of all single solutions. In the detail experiment, those thousands of solutions used in SCiM experiment were calculated in detail for the accurate proportions of all the components. According to the proportions, the absorbance curves were obtained. Any solution with a high concentration was diluted to the range in accordance with the Lambert-Beer law. And the absorbance curves of all the samples were calculated from the weighted sum of that of the original solutions. In this way, we can acquire enough spectrum data for ML in a short period of time. On the other hand, the practice is also an exploration of the automatic production of spectrum data.

Fig.2 shows the dataset for the measured 1,250 samples, including 1,250 SCiMs images (224 x 244 dpi) and the absorbance data in the wavelength range of 350-800 nm. To express this data, these images of the SCiMs are grouped into two 25-by-25-sized matrices (entries in Supplementary Table 3). The absorbance data are presented in the form of heat maps. The intensity of color in the image is proportional to the absorbance of the samples. At the same time, we selected randomly 8 samples to transform the related data into absorbance profiles. These obtained data will be provided as datasets for the next training of ML models.

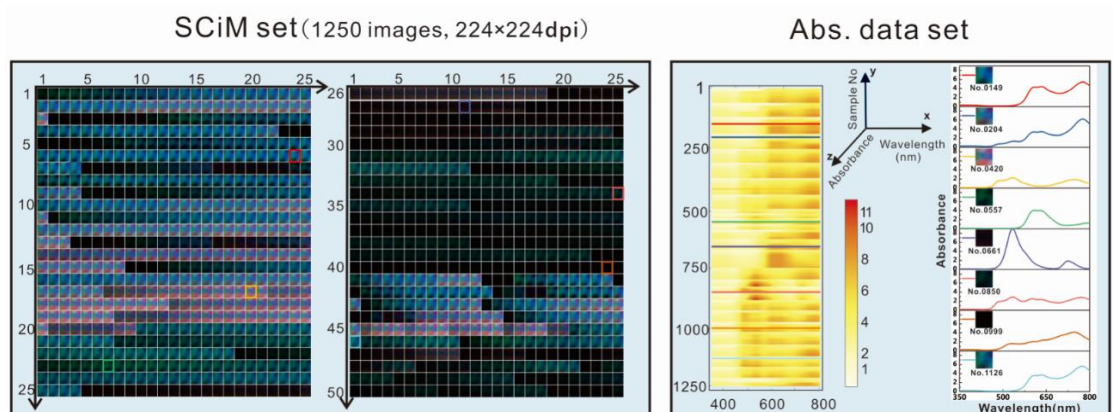

**Fig. 2. The obtained two data groups, SciM Set, and another of corresponding spectrum data measured by the spectrometer.** These 8 curves (in the rightside, as graphic examples) correspond to 8 given SCiMs (colored box in SCiM Set).

**Spectral prediction using CNN**
CNNs have the ability to extract image features, and deeper CNNs can extract more specific and complex features [16]. Whereas, for improving the CNN performance, the mere increment of network depth and width will bring about huge training parameters and over-fitting [17, 18]. For training SCiM dataset, we utilized and adjusted a sophisticated CNN model, GoogLeNet Inception v1 [19], which had a good performance in a known mission (the 2014 ImageNet Large Scale Visual Recognition Challenge) [20]. The key function blocks in the CNN, Inception Modules, will run concurrently multiple convolution and pooling operations on an input image. These structures are not only conducive to the formation of sparse connectivity (benefitting to simplify the training processes), but also to improve the computing performance of the network with its dense topology [19, 20]. Here, on one SCiM image, there are about $10^5$ pixels, and every pixel is an effective spectral feature points. So, compared with the images in some traditional dataset, for SCiMs training, we more focus on the effective extraction of the features as many as possible. The CNN of Inception v1 is a more appropriate candidate. Fig. 3 shows the working flow. The Inception v1 network (details see Supplementary Table 4) was used to extract the features from the image data, and to analysis the regression relationship between the SCiMs and the absorbance data. For adapting the spectrum data, the output of the CNN was modified into a linear layer with 451-dimension feature vectors. In detail, an input SCiM (224×224 dpi, RGB format) is processed by multiple layers of convolution, nonlinearity, and pooling operations, then give rise to a series of absorbance values (350 to 800 nm of wavelengths) at the output end.

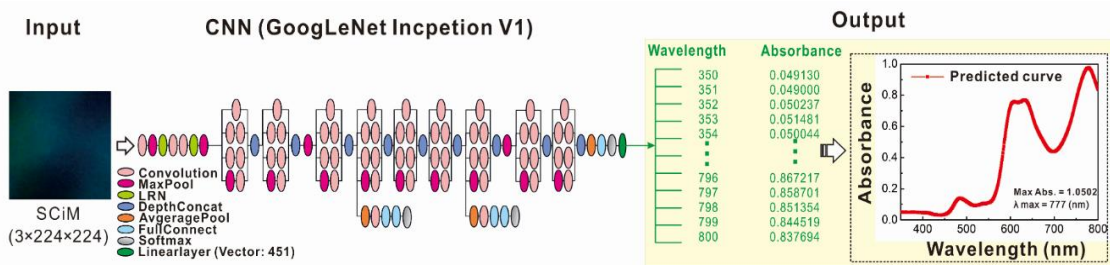

**Fig. 3. A CNN framework (A regression-based Google Inception v1) for the spectral prediction from SCiMs.** Data flow is from left to right. There are 22 main parameter layers for processing the input SCiMs, and the output layer containing 451 vectors (corresponding to the intervals in wavelengths of 350 to 800 nm), which will give the predicted absorbance curve of an input.

The predicted absorbances using CNN were exhibited in data maps (Fig. 4), containing all the experimental and predicting results on the training, validation and test sets. At the training stage, the collected dataset for the model learning was divided into 80% as a training set and 20% as a validation set. The absorbance values (divided into 451wavelengths) were measured for every sample, and served as data-tags for the related SCiMs. In other words, each SCiM as a CNN input would get 451 absorbance values as predicted outputs. For evaluating the trained CNN, a test set of new 138 SCiMs were predicted to inspect the performance. Besides, absolute errors between the experimental and the predicted values were calculated (right side of Fig. 4). Here, the prediction accuracy (represented by the absolute error) is negatively correlated with color depth on the maps.

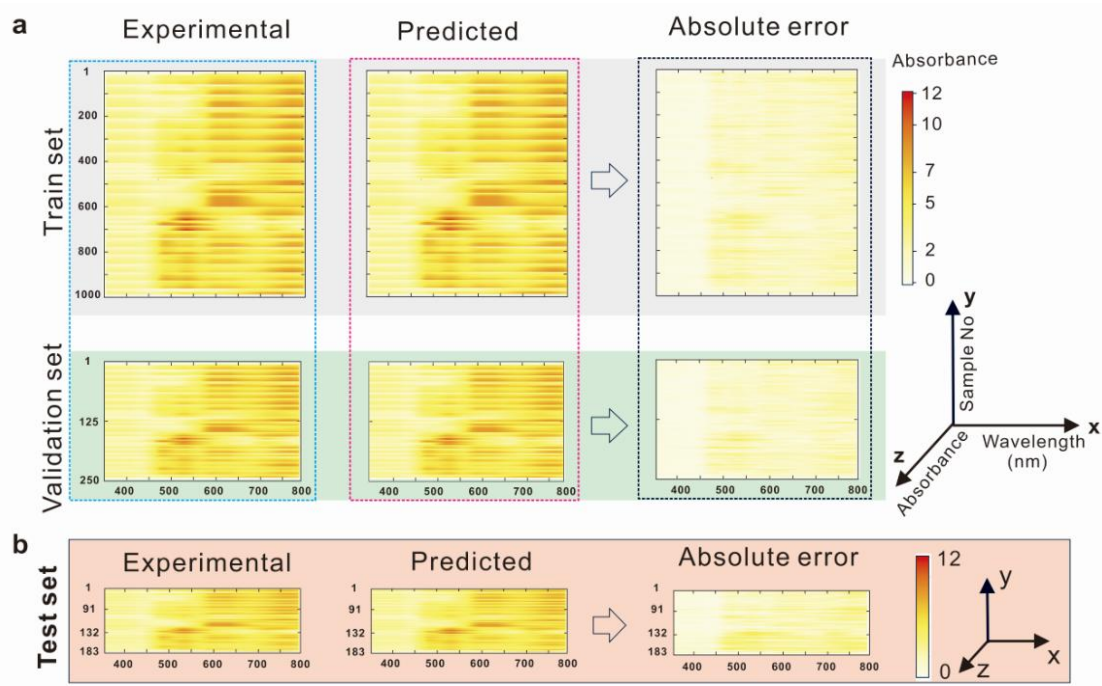

**Fig. 4. Application of machine learning to absorbance measurement.** (**a**) Data presentation of learning results (451 intervals in the wavelength range of 350 to 800 nm). The dataset of 1250 SCiMs as input was divided into an 1000/250 split of training and validation data. (**b**) Prediction results on the test set constituting of 138 SCiMs. Besides, the prediction performance was evaluated by the absolute errors between the experimental and the predicted values.

At the same time, we also evaluated the predictive accuracy of the CNN in other aspects. Prediction performance on all absorbance values (Fig. 5a) and the maximums of each sample (Fig. 5b) were evaluated on the test set (138 samples) and the $R^2$ value is 0.962 and 0.930 respectively. This means we obtained a satisfying result on the regression of the SCiMs without overfitting. Furthermore, the correlations between the experimental and predicted values were evaluated using Pearson correlation coefficient (PCC) [21]. The PCC is the covariance between two variables divided by the product of their standard deviation. The coefficients were always between from -1.0 to 1.0, closing to 0 was called no-correlative, and closing to 1 or -1 was called strongly positive or negative correlation [21]. For all the test samples (Fig. 5c), the mean of PCC was about 0.961, indicating a strong positive correlation between experimental and predicted absorbance. And the number (PCC > 0.9) reached up to 92%. And then the relative error rates (see in Fig. 5d) between the predicted and the experimental wavelengths of the maximum absorbance were also located, and their mean error of 3.281% was in a rational range. Just because the proposed measurement technology is so novel that it has many uncertained elements, we did not have enough time to optimize all of them. Therefore, such performance is still quite satisfactory. We can find that some aspects of the technology are well worth optimizing, such as the characteristics of light source, ECF patterns, samplling, CNN algorithms. Optimization of them will greatly improve the accuracy of the prediction. We estimate that this optimization process can take about a year or more in the capacity of our lab in the future. But its achievement can be expected to produce a device that is comparable in precision to conventional instruments.

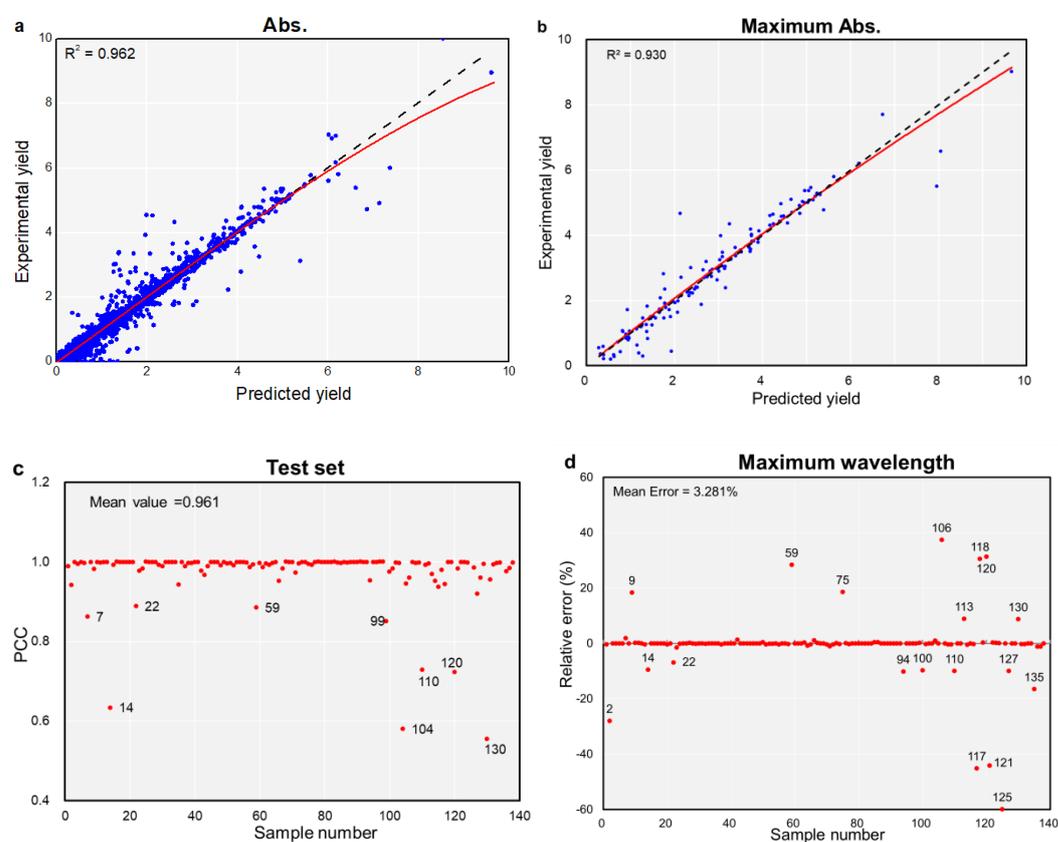

**Fig. 5. Performance of model on the test set (138 samples).** (**a**) is the prediction performance on all absorbance values (n = 62,238) and (**b**) is that on the maximums of each sample (n = 138). Superiority of the algorithm was evaluated by linear regression analysis: $R^2$ (coefficient of determination); dashed line, y = x line; solid line, Loess best-fit curves. (**c**) Pearson correlation coefficient (PCC) between the experimental and predicted values. The samples (PCC < 0.9) are marked separately, and their arithmetic mean was calculated. (**d**) Relative error rate between the predicted and experimental maximum wavelength (λ, nm). Those samples (error rate > 5%) are marked separately.

Additionally, we selected randomly 10 samples to present the prediction performance of the UV-VIS absorption spectra (Fig. 6). It can be found that the experimental and predicted spectra present highly similar. These results demonstrate the elaborately-trained CNN model will undertake eligibly the prediction task of absorption spectra based on the SCiMs.

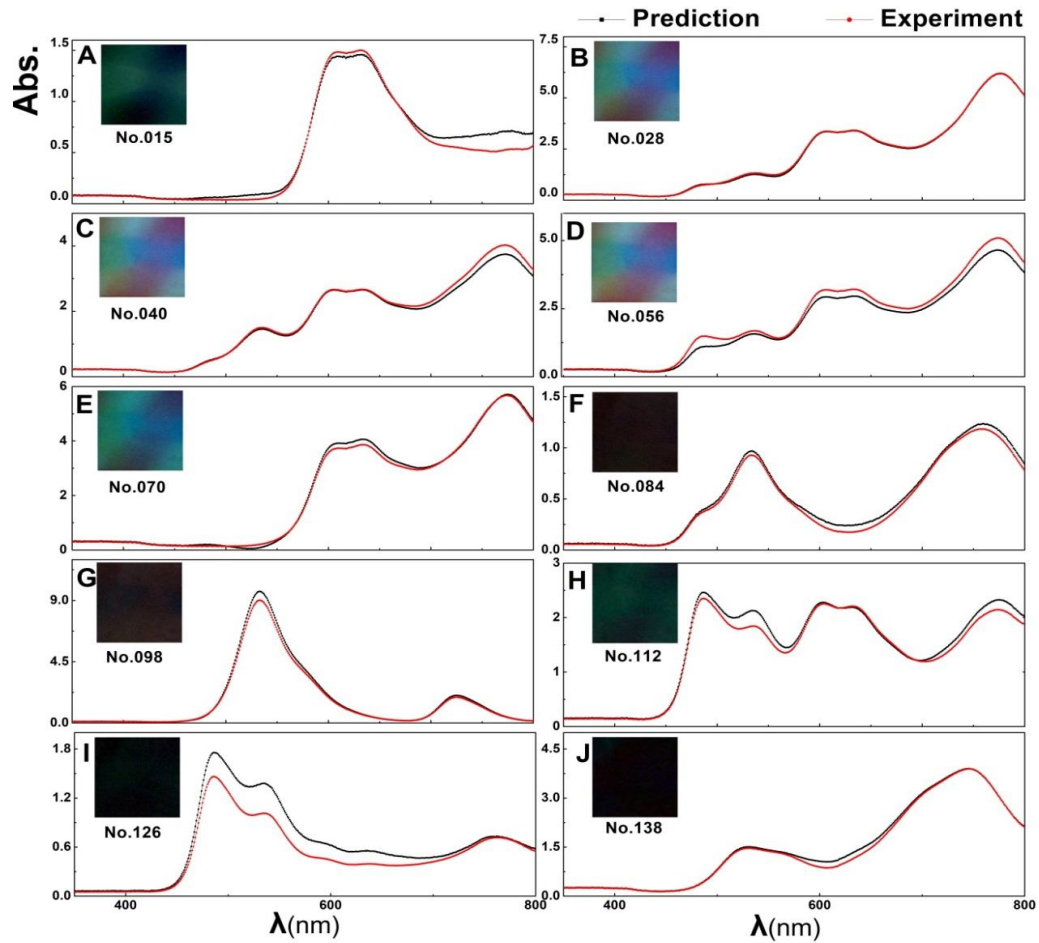

**Fig. 6.** UV-Vis absorbance curves (λ: 350 to 800 nm) of some samples selected randomly from the test set (138 samples).

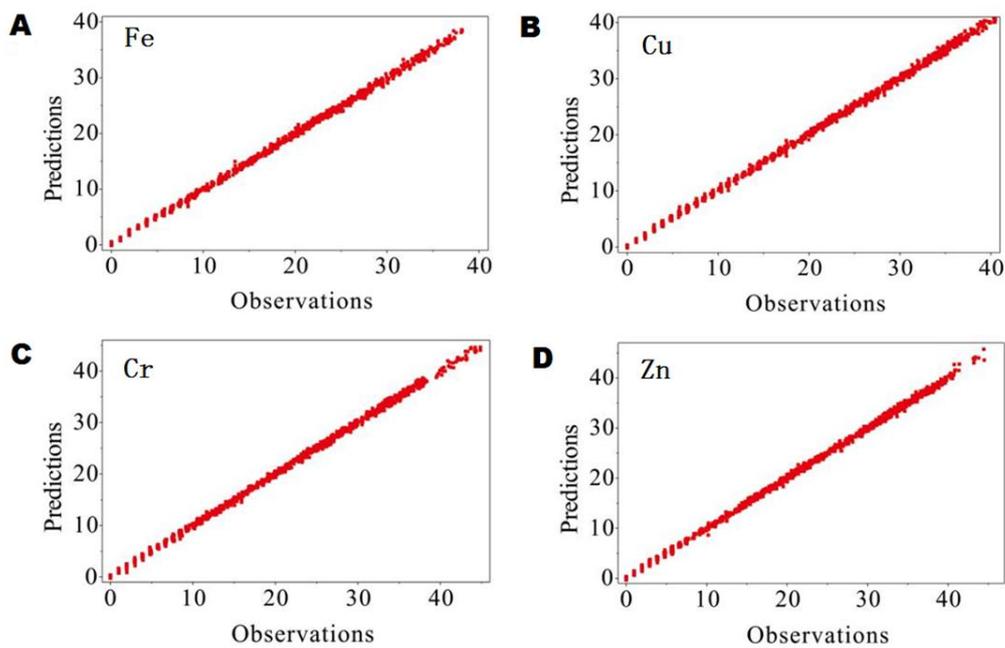

**Fig. 7.** Comparison of the actual concentration of various metals and the predicted concentration of the model. The unit of concentration is mg/L.

The above spectral technique can be used for spectral analysis of the composition of mixtures. A detailed description has been found in the document of a patented technology (CN201810770093.4). Complex mixed solutions are the most common phenomena in nature, such as urine, amino acids, soil solutions, plant tissue fluids, beverages, and the like. Here we conducted a verification experiment for the spectral analysis of mixed heavy metal solutions. The solution selected contains various concentrations of copper, iron, zinc, chromium. By adding a chromogenic reagent, the metal acts more strongly on the spectral changes. In this experiment, the composite indicator is a combination of aluminum reagent, naphthol green B, chrome black T, nitroso red salt, xylenol orange, and chrome azurol. A total of 1175 SciMs of the heavy metal mixed solution was collected through the constructed spectral experiment platform. Finally, a deep learning model is used to make predictions. The result is shown in Fig. 7. It can be seen that the prediction accuracy is very high.

**Discussion**

We think about the internal relationship between the SCiM images and the spectral properties, such as the absorbanceof matter. This relationship may be extremely complex and cannot be accurately described in terms of some definite approaches. However, we know that the difference between a SciM image and the sample-blank image is caused entirely by the light absorption of the sample. For presenting and managing such a complex difference, a neural network constructed with millions of nodes is a perfect carrier seemingly. A vividly statement is that a complex network with sufficient depth and breadth can describe this complex relationship digitally. However, for training a complex network successfully, a large amount of training data is unavoidable (e.g. the data amount of this study is ~ $10^{11}$ data points, i.e., $10^8 \times 1388$ chips). Practically, the tech-nodes for introducing DL to natural science often lies in how to gather sufficient datasets.

Previous researchers are less likely to choose such technical strategy. The reason is that these data obtained from a complex light is still a mixture of information, and are difficult to extract accurate and effective messages from the mixture. For this "meaningless" exploration, few researchers are willing to think and innovate further. Therefore, this work must wait for the emergence and improvement of big data and deep learning before being noticed by some scientists.

In this work, absorbance curves were chosen as the target of DL prediction. It is worth mentioning that there are actually many other properties that are directly or indirectly related to the spectrum. In practice, after obtaining sufficient amount of spectral data and any related variable, researchers can establish the relationship between the two through the ML approach. For example, medical scientists obtain millions of spectral data about human urine, and also collect some corresponding health data. A trained CNN will fully understand the expression of spectrum on health. This will be very meaningful in the medical and health field. This technology will evolve into a carrier and implementation tool for a wide range of medical tests. For the collection of sufficient training data, a good integration of hospitals and the technology will promote the achievement in just a few months. In the technical field of water environment management, similar spectral techniques will be very powerful in the resolution and description of the composition of water samples. They can also enable researchers to easily explore the relationship between spectroscopy and water quality via ML. In the future, water monitors will shuffle off the cumbersome and complex, and be constructed in density monitoring networks. Improve researchers' understanding of environmental processes.

Here, the cases performed in the visible-light range aimed mainly at verifying the feasibility of the proposed spectral technology. Referring to more practical applications, there are a lot of objects and ideas on them. For example, mixed chromogenic reagent is one of attractive and fantastic cases. We can combine several kinds of chromogenic chemicals to form a universal reagent. After the reagent is applied to samples, it will excite a large amount of spectral information on SCiMs. With machine learning, these SCiMs will greatly facilitate our understanding of research issues. These analysis technologies will make simple laboratories also have a strong ability to explore complex objects. From the perspective of wavelength, in ultraviolet or infrared band, the strategy of complicated beams is also effectual.

For example, in the UV band, researchers select several materials having different UV-absorbing properties to fabricated ECF, and can also obtain SCiMs of UV. In this course, the images are collected with a UV camera, or via fluorescence- conversion imaging with the UV excitation.

In summary, we have created a new method for measuring the spectral properties of matters using machine learning, and initially set up a device to realize the idea. The most immediate advantage is that it can greatly simplify the structure of traditional spectroscopic instruments. And through a series of designed experiments in the range of visible light, we verified the feasibility of the strategy and its performance. Obviously, this strategy is also viable in other spectrum range of UV, IR, x-ray. Due to the importance of spectrometry in the exploration of natural sciences, this work has been very instructive for researchers in many disciplines including chemistry, biology, environment, medicine, and food. It is also an important way for artificial intelligence to impact on scientific research.

## Materials and Methods
### Setup of SCiM equipment
The internal structure of the SCiM equipment is shown in Fig. 1, which is mainly composed of a light source, an ECF, a sample cell, and a detector. Among them, the primary light source is a combination of a 3-watt iodine tungsten lamp and a 0.25-watt white LED. A group of frosted glass is prepared from some transparent slide of quartz glass (thickness of 2 mm). The two windows of the sample cell are also made of quartz glass. The distance inside the two windows is 1 cm for controlling the liquid thickness. A CCD camera (T500, Sony) was used to detect and record the residual light. ECF was manufactured by the ink-jet printing (IJP) in a professional printing company. At first, a template (see in Supplementary Fig. 1) was designed as a rectangle pattern (1500×1500 dpi) composed of 3 color layers (cyan, magenta, and yellow). Then the template was printed on a transparent PET plate (size: 30×30 mm). After dried thoroughly, the printed plate was sealed on the surface of a quartz slide.

### Collection of SCiMs
We collected a total of 1,250 samples as the data for machine learning training. These samples were evenly distributed in a sample space consisting of all the mixtures of eight colored compounds. The space is also the definitional domain of the CNN training next. These 8 colored solutions (Supplementary Table 1) were prepared from the chemical reagents at analytical grade (absolute content ≥ 99%, purchased from Aladdin Sigma-Aldrich). The 1250 solutions have various given proportions and concentrations.

The SCiM images of these samples were collected by the above SCiM equipment. After running of 30 minutes for radiation stability, the SCiM equipment began to carry out the measurement of all the sample solutions in turn. In addition, distillated water as the blank controller was also conducted in on the same condation. The recorded photos by the CCD were the raw images of SCiMs. The photographing parameters are listed in Supplementary Table 3. The obtained images were standardized. First, the original images were uniformly adjusted to a size of 500 × 500 dpi. Then, all them were subtracted the blank image (the image obtained from a sample of pure water). This step is implemented in Matlab by image subtraction to gain the final SCiMs. It should be noted that the difference images are just derived from light absorption and have a closer relationship with absorbance curves.

Besides, after conducting CNN learning training, we collected the SCiMs (shown in Supplementary Fig. 2) for 138 new samples again. These samples were different at the constituent ratios but were similar at the measuring conditions. These samples were used to test the performance of the trained CNN.

### Collection of spectral data
In this work, the absorbance values ($\lambda$: 350 to 800 nm) was as the investigated variable of spectra phenomena. We scanned and measured the values of all one-component solutions of eight colored compounds by a UV–vis spectrometer (UV-2450, Shimadzu, Japan). Referring to the ratio of the above 1,250 samples in the work of SCiMs collection, the absorbance curves were calculated following their mixing ratios. All the experimental (and predicted) spectral data were shown in Extended Data Table S1 to 8.

**CNN Learning**
In this work, Google's Inception v1 CNN architecture was imported as a structure module during the training process. The major layers of the network included multiple convolutions, ReLU nonlinearity, max pooling, fully connected layers, ect. (details see in Supplementary Table 4). But compared with the previous architecture, we removed the final classification layer from the network, and added a linearlayer having 451 feature vectors. The modified CNN is trained using backpropagation that can match the prediction value and factual value to the closest degree. All layers of the network are fine-tuned using the learning rate of 0.001, max training grounds of 500 and batch size of 24. And the trained framework is capable to avoid over-fitting problem by introducing dropout layers (reducing the dimension via eliminating the relevant parameters) and regularization layers.

When CNN training, the 1250 SCiM images were divided into two parts, training set (1000 images) and validation set (250 images). During training we resize all images to 224×224 pixels to make it compatible with the original dimensions of the Inception v1 network architecture. In detail, when a SCiM of test sample is taken as the input, the network processes the images through multiple hidden layers in the CNN model, and then outputs the predicting outcomes by a fully connected layer with 451 feature vectors. The errors between experimental and predicted values of these vectors were used to calculate the loss function and to make absorbance predictions in the training and test stages, respectively. That is, given an input SCiM of a sample, the CNN model will give rise to a series of absorbance values.

**Data and materials availability**
The data needed to evaluate the conclusions in the paper are present in the paper and/or the Supplementary Materials. All the SCiM images and the supplementary can be downloaded from an URL, http://yunpan.snnu.edu.cn:80/#/link/4A9CBD96910F44B1949D36FEC1A1D272, with a fatch code: naPK. All the data that support the results of this study are available from the corresponding author.

**Acknowledgments**
This work is supported by the National Natural Science Foundation of China (No.50309011) and the Scientific Research Foundation for the Returned Overseas Chinese Scholars (08501041585). We thank to JiaBo Guo, Yao Wang, Yun Lee and Xin Yan for their work in preparing the materials in the early stage. And finally, we want to salute to the great physicist, Stephen William Hawking.

**Author contributions**
J.C.L. designed the study and analyzed the data; Q.N.D., S.F.B., and R.L. carried out experiments of preparation of ECF, Y.C. L., R.X.T., J.Y.C., G.D.M. and J.H.G did the experiments on collection of spectral data; J.C.L., Q.N.D and H.Q. L. conceptualized and trained the algorithms. Z.Y. X coordinated the project. All authors contributed to manuscript preparation.

**Competing interests**
The authors declare no competing interests.